\documentclass[journal,article,submit,moreauthors,dvipdfm,12pt,a4paper]{mdpi} 
\usepackage{amsmath}
\usepackage{amstext}
\usepackage{amsthm}
\usepackage{amssymb}
\usepackage{graphicx}
\usepackage{mathrsfs}
\usepackage{caption3}
\usepackage{caption}

\newcommand{\be}{\begin{equation}}
\newcommand{\ee}{\end{equation}}
\newcommand{\ba}{\begin{eqnarray}}
\newcommand{\ea}{\end{eqnarray}}
\newcommand{\bc}{\begin{center}}
\newcommand{\ec}{\end{center}}

\newcommand{\bay}{\begin{array}{rcl}}
\newcommand{\eay}{\end{array}}

\def\ie{{i.e. }}
\def\lp{\ell_{\rm Pl}}
\def\LH{\ell_{\rm H}}

\def\tr{t_{\rm tr}}
\def\mp{m_{\rm Pl}}
\def\tp{t_{\rm Pl}}
\def\OM{\Omega_{\rm M}}
\def\OL{\Omega_{\Lambda}}

\def\OLS{\Omega_\Lambda^\ast}

\def\barg{\bar{G}}
\def\lat{\lambda_{\rm T}}

\def\kat{k_{\rm T}}
\def\gat{g_{\rm T}}
\def\hT{H_{\rm T}}
\def\h0{H_{0}}
\def\aT{a_{\rm T}}

\def\tT{t_{\rm T}}

\def\calp{\widetilde{\mathscr P}}
\def\clp{{\mathscr P}}
\def\clm{{\cal M}}
\def\UU{{\cal U}}
\def\ssc{S_{\rm c}}
\lastpage{x}
\doinum{10.3390/------}
\pubvolume{xx}
\pubyear{2010}
\history{Received: xx / Accepted: xx / Published: xx}

\Title{Entropy Production during Asymptotically Safe Inflation\footnote{Invited contribution to the special issue of {\it Entropy}
on entropy in quantum gravity.}}

\Author{A. Bonanno$^{1,2}$ and M. Reuter$^3$ }
\address{$^1$INAF-Catania Astrophysical Observatory,Via S.Sofia 78, 95123 Catania, Italy. \\
$^2$INFN, Via S.Sofia 64, 95123 Catania, Italy.\\
$^3$Institute of Physics, University of Mainz, Staudingerweg 7, D-55099 Mainz, Germany}

\corres{ Alfio Bonanno, abo@oact.inaf.it}

\abstract{
The Asymptotic Safety scenario predicts that the deep ultraviolet of Quantum Einstein Gravity is governed by 
a nontrivial renormalization group fixed point. Analyzing its implications for cosmology using 
renormalization group improved Einstein equations we find that it can give rise to a phase of inflationary 
expansion in the early Universe. Inflation is a pure quantum effect here and requires no inflaton field. It is driven by the
cosmological constant and ends automatically when the renormalization group evolution has reduced the vacuum energy 
to the level of the matter energy density.  The quantum gravity effects also provide a natural  mechanism for the generation of entropy.
It could easily account for the entire entropy of the present Universe in the massless sector.
}


\begin{document}

\section{Introduction}
After the introduction of the effective average action and its functional renormalization group equation for gravity
\cite{mr} detailed investigations of the nonperturbative renormalization
group (RG) behavior of
Quantum Einstein Gravity (QEG) have become possible
\cite{mr,percadou,oliver1,frank1,oliver2,oliver3,oliver4,frank2,souma,perper1,codello,
litimgrav,prop,colombia,essential}.
The exact RG equation underlying this approach defines a Wilsonian
RG flow on a theory space which consists of all diffeomorphism invariant
functionals of the metric $g_{\mu\nu}$. The approach   turned out
to be an ideal setting for investigating the Asymptotic Safety scenario in gravity
\cite{wein,livrev} and, in fact, substantial evidence was found for the
nonperturbative renormalizability of Quantum Einstein Gravity.
The theory emerging  from this construction (``QEG")
is not a quantization of classical general relativity. Instead, its bare action
corresponds to a nontrivial fixed point of the RG flow and is
a {\it prediction} therefore.
The effective average action  \cite{mr,avact,ym,avactrev}
has crucial advantages as compared to other continuum implementations of the
Wilson RG, in particular it is closely related to the standard
effective action and defines a family of effective field theories
$\{ \Gamma_k[g_{\mu\nu}], 0 \leq k < \infty \}$ labeled by the coarse graining scale $k$.
The latter property opens the door to a rather direct extraction of physical
information from the RG flow, at least in single-scale cases: If the physical process
or phenomenon under consideration involves only a single typical momentum
scale $p_0$ it can be described by a tree-level evaluation of $\Gamma_k[g_{\mu\nu}]$, with $k=p_0$.
The precision which can be achieved by this effective field theory description
depends on the size of the fluctuations relative to the mean values. If they are large, or if more than
one scale is involved, it might be necessary to go  beyond the tree analysis.
The RG flow of the effective average action, obtained by different truncations of theory space,
has been the basis of various investigations of ``RG improved" black hole and cosmological
spacetimes \cite{bh,erik1,cosmo1,cosmofrank}.
We shall discuss some aspects of this method below.

The purpose of this article is to review the main features of renormalization group improved cosmologies 
based upon a RG trajectory of QEG with realistic parameter values. As a direct consequence
of the nontrivial RG fixed point which underlies Asymptotic Safety the early  Universe is found to 
undergo a phase of adiabatic inflationary expansion; it is a pure quantum effect and requires no inflaton field.
Furthermore, we shall see that the quantum gravity effects provide a novel mechanism for the generation of entropy;
in fact, they easily could account for the entire entropy of the present Universe in the massless sector. 

Our presentation follow \cite{ourfluc} and \cite{pune} to which the reader is referred for further details.
A related investigation of  ``asymptotically safe inflation", using different methods, has been performed by 
S. Weinberg \cite{weinf}.

\section{Entropy and the renormalization group}

A special class of RG trajectories obtained from QEG in the Einstein-Hilbert approximation
\cite{mr}, namely those of the ``Type IIIa'' \cite{frank1}, possess all the qualitative
properties one would expect from the  RG trajectory describing gravitational phenomena
in the real Universe we live in. In particular they can have a long classical regime and a small,
positive cosmological constant in the infrared (IR). Determining its parameters from observations,
one finds \cite{ourfluc} that,  according to this particular QEG trajectory, the running cosmological
constant $\Lambda(k)$ changes by about 120 orders of magnitude between $k$-values of the order
of the Planck mass and macroscopic scales, while the running Newton constant $G(k)$ has no
strong $k$-dependence in this regime. For $k> \mp$, the non-Gaussian fixed point (NGFP)
which is responsible for the Asymptotic Safety of QEG controls their scale dependence. In the deep
ultraviolet $(k\rightarrow \infty)$, $\Lambda(k)$ diverges and $G(k)$ approaches zero.

An immediate question is whether there is any experimental or observational 
evidence that would  hint at this enormous scale dependence of the gravitational parameters,
the cosmological constant in particular. Clearly the natural place to search for such phenomena
is cosmology. Even though it is always difficult to give a precise physical interpretation
to the RG scale $k$ is is fairly certain that any sensible identification of $k$ in terms
of cosmological quantities will lead to a $k$ which decreases during the expansion
of the Universe. As a consequence, $\Lambda(k)$ will also decrease as the Universe expands.
Already the purely qualitative assumption of a {\it positive} and {\it decreasing}
cosmological constant supplies an interesting hint as to which 
phenomena might reflect a possible $\Lambda$-running. 
\begin{figure}[h]
\begin{center}$
\begin{array}{cc}
\includegraphics[width=2.5in]{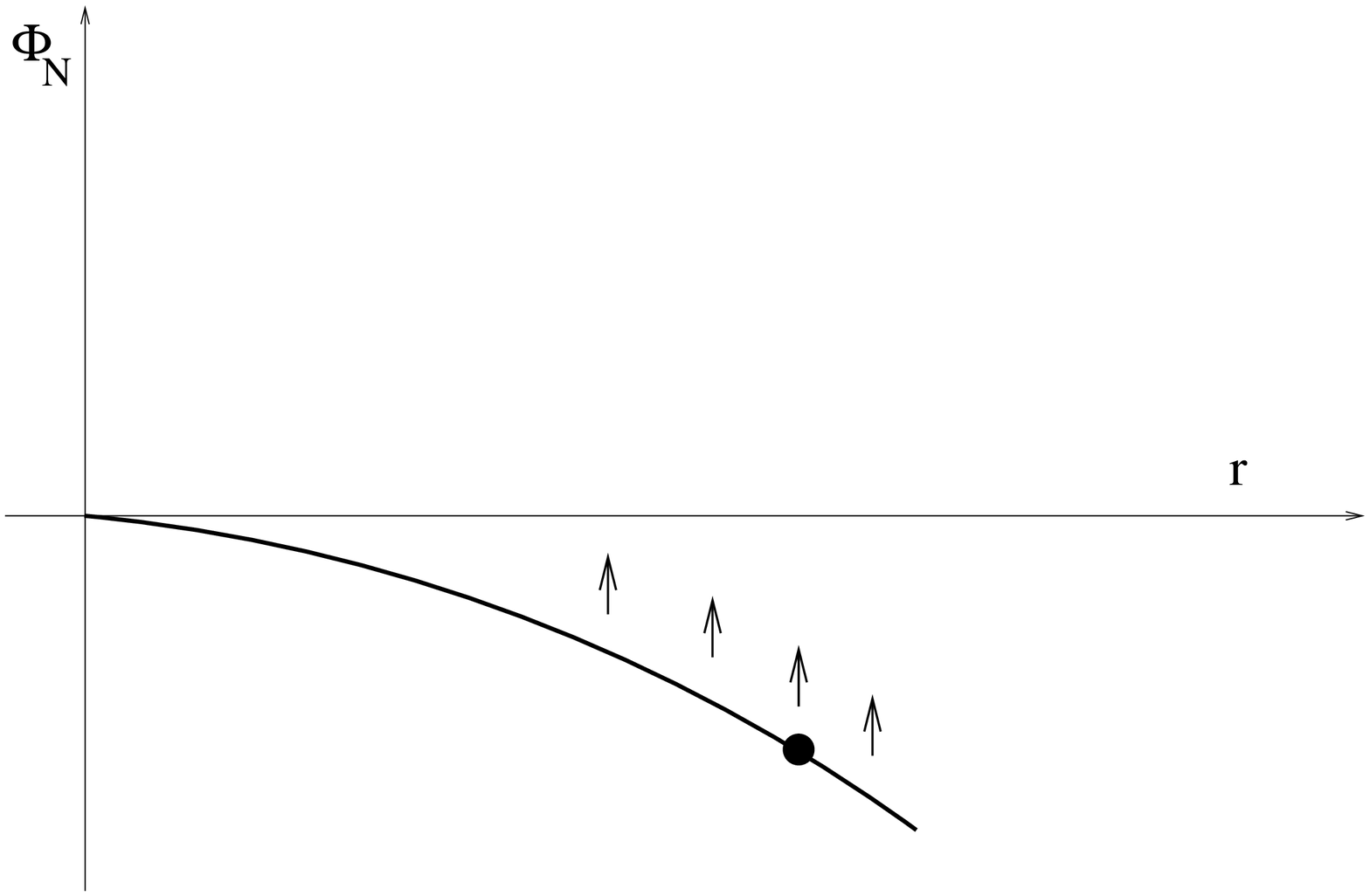}&
\includegraphics[width=3.5in]{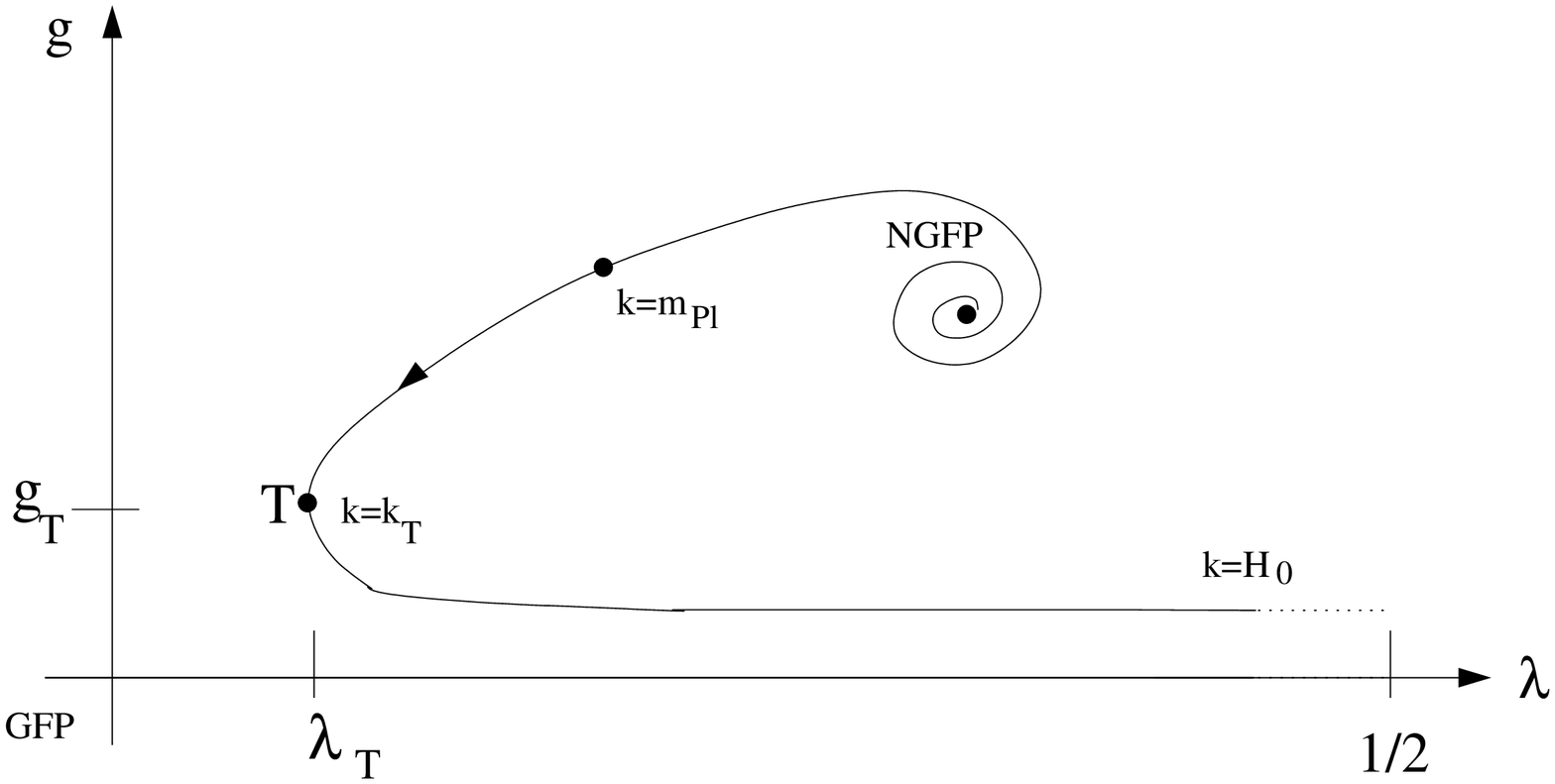}
\end{array}$
\end{center}
\caption{ The left panel shows the quasi-Newtonian potential corresponding to de Sitter space.
The curve moves upward as the cosmological constant decreases. 
On the right panel the ``realistic" RG trajectory discussed in Section 5 is shown.} 
\label{fig1}
\end{figure}

To make the argument as simple as possible, let us first consider a Universe without matter,
but with a positive $\Lambda$. Assuming maximal symmetry, this is nothing but  de Sitter
space, of course. In static coordinates its metric is 
\[
ds^2 = -(1+2\Phi_{\rm N}(r) ) dt^2+ (1+2\Phi_{\rm N}(r))^{-1}dr^2 +
r^2 (d\theta^2 +\sin^2 \theta d\phi^2)
\]
with 
\[
\Phi_{\rm N}(r) = -\frac{1}{6}\; \Lambda\; r^2.
\]
In the weak field and slow motion limit $\Phi_{\rm N}$ has the interpretation
of a Newtonian potential, with a correspondingly simple physical interpretation.
The left panel of Fig.1 shows $\Phi_{\rm N}$ as a function of $r$; for $\Lambda >0$ it is 
an upside-down parabola. Point particles in this spacetime, symbolized by the 
black dot in Fig.1, ``roll down the hill'' and are rapidly driven away
from the origin and from any other particle. Now assume that the magnitude of $|\Lambda|$
is slowly (``adiabatically'') decreased. This will cause the potential $\Phi_{\rm N}(r)$ 
to move upward as a whole, its slope decreases. So the change in $\Lambda$ increases the particle's
potential energy.  This is the simplest way of understanding that a {\it positive decreasing}
cosmological constant has the effect of ``pumping'' energy into the matter degrees of freedom.
More realistically one will describe the matter system in a hydrodynamics or quantum field
theory language and one will include its backreaction onto the metric. But the basic conclusion, 
namely that a slow decrease of a positive $\Lambda$ transfers energy into the matter system, will
remain true. 

We are thus led to  suspect that, because of the decreasing cosmological constant, 
there is a continuous inflow of energy into the cosmological fluid contained in an 
expanding Universe. It will ``heat up'' the fluid or, more exactly, lead to a slower decrease
of the temperature than in standard cosmology. Furthermore, by elementary thermodynamics, it will 
{\it increase} the entropy of the fluid. If during the time $dt$ an amount of heat
$d Q>0$ is transferred into a volume $V$ at the temperature $T$ the entropy changes by an amount $dS=dQ/T>0$.
To be as conservative (i.e., close to standard cosmology) as possible, we assume that this process
is reversible. If not, $dS$ is even larger.

In standard Friedmann-Robertson-Walker (FRW) cosmology the expansion is adiabatic,
the entropy (within a comoving volume) is constant. It has always been somewhat puzzling therefore
where the huge amount of entropy contained in the present Universe comes from. 
Presumably it is dominated by the CMBR photons which  contribute an amount
of about $10^{88}$ to the entropy within the present Hubble sphere. (We use units such that
$k_{\rm B}=1$. ) In fact, if it is really true that no entropy is produced during the 
expansion then the Universe would have had an entropy of at least $10^{88}$ immediately 
after the initial singularity which for various reasons seems quite unnatural.
In scenarios which invoke a ``tunneling from nothing'', for instance,
spacetime was ``born'' in a pure quantum state, so the very early Universe is expected to have 
essentially no entropy. Usually it is argued that the entropy present today is the result
of some sort of ``coarse graining'' which, however, typically is not considered an active part of the
cosmological dynamics in the sense that it would have an impact on the 
time evolution of the metric, say.

Following \cite{ourfluc}  we shall argue that in principle the entire entropy of the massless fields in the present
universe can be understood as arising from the mechanism described above.  
If energy can be exchanged freely between
the cosmological constant and the matter degrees of freedom, the entropy observed today 
is obtained precisely if the initial entropy at the ``big bang'' vanishes.
The assumption that the matter system must allow for an unhindered energy exchange with
$\Lambda$ is essential, see  refs. \cite{cosmo1,ourfluc}.

We shall model the matter in the early Universe by a gas with $n_{\rm b}$ 
bosonic and $n_{\rm f}$ fermionic massless degrees of freedom, all at  the same
temperature. {\it In equilibrium} its energy density, pressure, and entropy density
are given by the usual relations ($n_{\rm eff}=n_{\rm b}+\frac{7}{8}n_{\rm f}$)
\begin{subequations}\label{1.3}
\ba\label{1.3a}
&&\rho = 3\; p = \frac{\pi^2}{30} \; n_{\rm eff} \; T^4\\[2mm] 
&&s = \frac{2\pi^2}{45}\; n_{\rm eff} \; T^3\label{1.3b}\\[2mm]
&&\text{ \hspace{-7.45cm}  so that in terms of $U\equiv \rho \; V$ and $S\equiv s\; V$,} \nonumber\\[2mm]
&&T\; dS = dU + p\; dV\label{1.3c}
\ea
\end{subequations}
In an out-of-equilibrium process of entropy generation the question arises 
how the various thermodynamical quantities are related then. To be as conservative as possible,
we make the assumption that the irreversible inflow of energy
destroys thermal equilibrium as little as possible in the sense that the equilibrium
relation (\ref{1.3}) continue to be (approximately) valid.

This kind of thermodynamics in an FRW-type cosmology with a decaying cosmological constant
has been analyzed in detail by Lima \cite{lima1}, see also \cite{lima2}. It was shown that if
the process of matter creation $\Lambda(t)$ gives rise to is such that the specific entropy
per particle is constant, the relations of equilibrium thermodynamics are preserved.
This means that no finite thermalization time is required since the particles originating from 
the decaying vacuum are created in equilibrium with the already existing ones. Under these
conditions it is also possible to derive a generalized black body spectrum which is
conserved under time evolution. Such minimally non-adiabatic processes
were termed ``adiabatic'' (with the quotation marks) in refs. \cite{lima1,lima2}.

\section{Asymptotically Safe Inflation}

There is another, more direct potential consequence of a decreasing positive cosmological
constant which we shall also explore here, namely a period of automatic
inflation during the very first stages of the cosmological evolution. 
In the very early Universe the RG running of the gravitational parameters is governed by the non-Gaussian
RG fixed point which is at the heart of the Asymptotic Safety scenario.
The inflationary phase we are going to describe is a rather direct consequence of the huge cosmological 
constant which the NGFP enforces during  the epoch governed by the asymptotic scaling regime of the 
renormalization group.

It is not surprising,
of course, that a positive $\Lambda$ can cause an accelerated expansion, but in the
classical context the problem with a $\Lambda$-driven inflation is that it would never
terminate once it has started. In popular models of scalar driven inflation 
this problem is circumvented by designing the inflaton potential in such a way that it
gives rise to a vanishing vacuum energy after a period of ``slow roll''. 

As we shall see   generic RG cosmologies based upon the QEG trajectories  
have an era of $\Lambda$-driven inflation immediately after the big bang which ends automatically as a 
consequence of the RG running of $\Lambda(k)$. Once the scale $k$ drops significantly below 
$\mp$, the accelerated expansion ends because the vacuum energy density $\rho_\Lambda$ is already
too small to compete with the matter density. Clearly this is a very attractive scenario:
{\it neither to trigger inflation nor to stop it one needs any ad hoc ingredients such
as an inflaton field or a special potential}. It suffices to include the leading quantum
effects in the gravity + matter system.
Furthermore, asymptotic safety offers a natural mechanism for the quantum mechanical generation
of primordial density perturbations, the seeds of cosmological structure formations. 

In the following we review a concrete investigation along these
lines.  For further details we  refer to \cite{cosmo1,ourfluc}.

\section{The improved Einstein equations}
The computational setting of our investigation \cite{ourfluc} are the RG
improved Einstein equations: By means of a suitable cutoff identification $k=k(t)$
we turn the scale dependence of $G(k)$ and $\Lambda(k)$ 
into a time dependence, and then substitute the resulting $G(t)\equiv G(k(t))$
and $\Lambda(t)\equiv \Lambda(k(t))$ into the Einstein equations
$G_{\mu\nu}=-\Lambda(t)g_{\mu\nu}+8\pi G(t) T_{\mu\nu}$. 
We specialize $g_{\mu\nu}$
to describe a spatially flat $(K=0)$ Robertson-Walker metric with scale factor $a(t)$,
and we take ${T_\mu}^\nu = {\rm diag}[-\rho,p,p,p]$ to be the energy momentum
tensor of an ideal fluid with equation of state $p=w\rho$ where $w>-1$ is constant.
Then the improved Einstein equation boils down 
to the modified Friedmann equation and a continuity equation:
\begin{subequations}
\ba\label{2.5a}
&&H^2 = \frac{8\pi}{3} G(t) \; \rho + \frac{1}{3}\Lambda(t)\\[2mm]
&&\dot\rho+3H(\rho+p)=-\frac{\dot{\Lambda}+8\pi \; \rho \; \dot{G}}{8\pi \; G}\label{2.5b}
\ea
\end{subequations}
The modified continuity equation (\ref{2.5b}) is the integrability condition for the 
improved Einstein equation implied by Bianchi's identity, 
$D^\mu [-\Lambda(t) g_{\mu\nu} + 8\pi G(t) T_{\mu\nu}]=0$. 
It describes the energy exchange 
between the matter and gravitational degrees of freedom (geometry).
For later use let us note that upon
defining the critical density
$\rho_{\rm crit}(t)\equiv {3 \; H(t)^2}/{8\pi \; G(t)}$
and the relative densities $\Omega_{\rm M}\equiv \rho/\rho_{\rm crit}$ and 
$\Omega_\Lambda=\rho_\Lambda/\rho_{\rm crit}$ the modified Friedmann equation (\ref{2.5a})
can be written as
$\OM(t) +\OL(t) = 1$.

We shall obtain $G(k)$ and $\Lambda(k)$ by solving the flow equation in the Einstein-Hilbert
truncation with a sharp cutoff \cite{mr,frank1}. It is formulated in terms of the dimensionless
Newton and cosmological constant, respectively:
$g(k)\equiv k^2 \; G(k)$, $\lambda(k) = \Lambda(k)/k^2$.
We then construct quantum corrected cosmologies by (numerically) solving the RG improved evolution equations.
We shall employ the cutoff identification
\be\label{2.12}
k(t) =\xi H (t)
\ee
where $\xi$ is a fixed positive constant of order unity.
This is a natural choice since in a Robertson-Walker geometry
the Hubble parameter measures the curvature of spacetime; its inverse $H^{-1}$ defines 
the size of the ``Einstein elevator''. Thus we have
\be\label{2.13}
G(t) = \frac{g(\xi H(t))}{\xi^2 \; H(t)^2}, 
\;\;\;\;\;\;\;\;\;\; \Lambda(t) = \xi^2 \; H(t)^2 \; \lambda(\xi H(t))
\ee
One can prove  that all  solutions of the coupled system of differential
equations (\ref{2.5a}, \ref{2.5b}) can be obtained by means of the following algorithm: 

{\it Let $ \Big ( g(k),\lambda(k) \Big )$ be a prescribed RG trajectory 
and $H(t)$ a solution of }
\be\label{2.14}
\dot{H}(t)= -\frac{1}{2}(3+3w)H(t)^2\Big [ 1-\frac{1}{3} \; \xi^2 \; \lambda(\xi H(t)) \Big]
\ee
{\it Let $\rho(t)$ be defined in terms of this solution by }
\be\label{2.155}
\rho(t) = \frac{3 \; \xi^2}{8\pi\; g(\xi H(t))} 
\Big [ 1-\frac{1}{3}\; \xi^2 \;  \lambda(\xi H(t)) \Big ] \, H(t)^4
\ee
{\it Then the pair $\Big( H(t),\rho(t) \Big )$ 
is a solution of the system} (\ref{2.5a}), (\ref{2.5b})
{\it for the time dependence of $G$ and $\Lambda$ 
given by} (\ref{2.13}) {\it and the equation of state $p=w\rho$, provided $H(t)\not =0$. }

\section{RG trajectory with realistic parameter values}

Before we start solving the modified field equations
let us briefly review how the type IIIa trajectories of the Einstein-Hilbert truncation
can be matched against the observational data \cite{bh}. This analysis is fairly robust 
and clearcut; it does not involve the NGFP. All that is needed is the RG flow linearized 
about the Gaussian fixed point (GFP) which is located at $g=\lambda=0$. In its vicinity one
has \cite{mr}
$\Lambda(k) = \Lambda_0 + \nu \; \barg \, k^4+\cdots$ and $G(k) =\barg+\cdots$.
Or, in terms of the dimensionless couplings, 
$\lambda(k) = \Lambda_0/k^2 + \nu \; \barg \, k^2+\cdots$, $g(k) =\barg \; k^2+\cdots$.
In the linear  regime of the GFP, $\Lambda$ displays a running $\propto k^4$
and $G$ is approximately constant. Here $\nu$ is a positive 
constant of order unity \cite{mr,frank1},  
$\nu \equiv \frac{1}{4\pi} \Phi^{1}_{2}(0)$. These equations are valid if $\lambda(k) \ll 1$ and $g(k)\ll 1$. They
describe a 2-parameter family of RG trajectories labeled by the pair $(\Lambda_0, \barg)$.
It will prove convenient to use an alternative labeling $(\lat, \kat)$ with
$\lat \equiv (4  \nu  \Lambda_0 \barg)^{1/2}$ and 
$\kat \equiv   ( {\Lambda_0}/{\nu  \barg}  )^{1/4} $.
The old labels are expressed in terms of the new ones as
$\Lambda_0 = \frac{1}{2} \lat \; \kat^2$ and $\barg = {\lat}/{2\,\nu\,\kat^2}$.
It is furthermore convenient to introduce the abbreviation
$\gat\equiv {\lat}/{2\,\nu}$.
When parameterized by the pair $(\lat,\kat)$ the trajectories assume the form
\ba\label{1.16}
&&\Lambda(k) = \frac{1}{2} \; \lat\; \kat^2 \; \Big [ 1+(k/\kat)^4\Big ]
\equiv \Lambda_0 \Big [ 1+(k/\kat)^4 \Big ]\\[2mm]
&&G(k) = \frac{\lat}{2\,\nu\,\kat^2}\equiv \frac{\gat}{\kat^2}\nonumber
\ea
or, in dimensionless form, 
\be\label{1.17}
\lambda(k) = \frac{1}{2} \; \lat \Big  [  \Big (\frac{\kat}{k} \Big)^2+ 
 \Big ( \frac{k}{\kat}  \Big)^2 \Big ], \;\;\;\;\;\;\;\;\;\;  g(k) = \gat \,  \Big( \frac{k}{\kat}\Big  )^2
\ee
As for the interpretation of the new variables, 
it is clear that $\lat \equiv \lambda(k\equiv \kat)$ and $\gat\equiv g(k=\kat)$,
while $\kat$ is the scale at which $\beta_\lambda$ (but not $\beta_g$) vanishes according
to the linearized running:
$\beta_\lambda(\kat)\equiv k{d \lambda(k)}/{dk}  |_{k=\kat} =0$.
Thus we see that $(\gat,\lat)$ are the coordinates of the turning point T
of the type IIIa trajectory considered, and $\kat$ is the scale at which it is
passed. It is convenient to refer the ``RG time'' $\tau$ to this scale:
$\tau(k) \equiv \ln (k/\kat)$.
Hence $\tau>0$ ($\tau<0$) corresponds to the ``UV regime'' (``IR regime'') where
$k>\kat$ ($k<\kat)$. 

Let us now hypothesize that, within a certain range of $k$-values, the RG trajectory
realized in Nature can be approximated by (\ref{1.17}). In order to determine
its parameters $(\Lambda_0, \barg)$ or $(\lat, \kat)$ we must perform a measurement
of $G$ and $\Lambda$. If we interpret the observed values
$G_{\rm observed} = \mp^{-2}$, $\mp\approx 1.2\times 10^{19} \, {\rm GeV}$, and 
$\Lambda_{\rm observed} = 3\,\Omega_{\Lambda 0}\,H_0^2\approx 10^{-120}\, \mp^2\nonumber$
as the running $G(k)$ and $\Lambda(k)$ evaluated at a scale $k\ll \kat$, then we get from
(\ref{1.16}) that $\Lambda_0 =\Lambda_{\rm observed} $ and 
$\barg = G_{\rm observed}$. Using the definitions of  $\lat$ and $\kat$ along with $\nu = O(1)$ this leads to the 
order-of-magnitude estimates
$\gat\approx \lat \approx 10^{-60}$ and $\kat\approx 10^{-30}\;\mp\approx (10^{-3} {\rm cm})^{-1}$.
Because of the tiny values of $\gat$ and $\lat$ the turning point lies in the linear regime of the GFP. 

Up to this point we discussed only that segment of the ``trajectory realized in Nature'' which lies
inside the linear regime of the GFP. The complete RG trajectory obtains by continuing
this segment with the flow equation both into the IR and into the UV, 
where it ultimately spirals into the 
NGFP. While the UV-continuation is possible within the Einstein-Hilbert
truncation, this approximation breaks down in the IR when $\lambda(k)$ approaches $1/2$.
Interestingly enough, this happens near $k=H_0$, the present Hubble scale.
The right panel of Fig.1 shows a schematic sketch of the complete trajectory 
on the $g$-$\lambda$--plane and Fig.2 displays the
resulting $k$-dependence of $G$ and $\Lambda$.

\section{Primordial entropy generation}
Let us return to the modified continuity equation (\ref{2.5b}). After multiplication by $a^3$
it reads
\be\label{3.1}
[\dot\rho + 3H(\rho +p)] \; a^3 = \calp(t)
\ee
where we defined
\be\label{3.2}
\calp\equiv -\Big ( \frac{\dot{\Lambda}+8 \pi\; \rho \; \dot{G}}{8\pi \; G} \Big ) a^3
\ee
Without assuming any particular equation of state Eq.(\ref{3.1}) can be
rewritten as 
\be\label{3.3}
\frac{d}{dt} (\rho a^3) +p\frac{d}{dt}(a^3) = \calp(t)
\ee
The interpretation of this equation is as follows. 
Let us consider a unit {\it coordinate}, 
i.e. comoving volume in the Robertson-Walker spacetime. Its corresponding  
{\it proper} volume is $V=a^3$
and its energy contents is $U=\rho a^3$. The rate of change of these quantities is subject to 
(\ref{3.3}): 
\be\label{3.4}
\frac{dU}{dt}+p\frac{dV}{dt}=\calp(t)
\ee
In classical cosmology where $\calp\equiv 0$ this equation together with the standard thermodynamic
relation $dU+pdV=TdS$ is used to conclude that the expansion of the Universe is adiabatic, \ie
the entropy inside a comoving volume does not change as the Universe expands, $dS/dt=0$.

Here and in the following we write $S\equiv s \, a^3$ for the entropy carried by the matter inside
a unit comoving volume and $s$ for the corresponding proper entropy density.

When $\Lambda$ and $G$ are time dependent, $\calp$ is nonzero and we interpret (\ref{3.4})
as describing the process of energy (or ``heat'') exchange between the scalar fields $\Lambda$ 
and $G$  and the ordinary matter. This interaction causes $S$ to change:
\be\label{3.5}
T\frac{dS}{dt}=T\frac{d}{dt}(s a^3)=\calp(t)
\ee
The actual rate of change of the comoving entropy is 
\be\label{3.6}
\frac{dS}{dt}=\frac{d}{dt}(s a^3)= \clp (t)
\ee
where 
\be\label{3.7}
\clp \equiv \calp /T
\ee 
If $T$ is known as a function of $t$ we can integrate (\ref{3.5}) to obtain $S=S(t)$. 
In the RG improved cosmologies the entropy production rate per comoving 
volume 
\be\label{3.7}
\clp(t) = - \Big [ \frac{ \dot\Lambda+8\pi \; \rho \; \dot G}{8\pi\;  G} \Big ] \frac{a^3}{T}
\ee
is nonzero because the gravitational ``constants''  $\Lambda$ and $G$ have acquired a time 
dependence. 

For a given solution to the coupled system of RG and cosmological 
equations it is sometimes more convenient to calculate $\clp (t)$ from the LHS of the 
modified continuity equation rather than its RHS (\ref{3.7}): 
\be\label{3.8}
[\dot\rho + 3H(\rho +p)] \frac{a^3}{T} = \clp(t)
\ee

If $S$ is to increase with time, by (\ref{3.7}), we need that $\dot\Lambda +8\pi \dot{G}<0$.
During most epochs of the RG improved cosmologies we have $\dot\Lambda \leq 0$ and
$\dot G\geq 0$. The decreasing $\Lambda$ and the increasing $G$ have antagonistic effects 
therefore. We shall see that in the physically realistic cases $\Lambda$ predominates so that
there is indeed a transfer of energy from the vacuum to the matter sector rather than
vice versa.

Clearly we can convert the heat exchanged, $TdS$, to an entropy change only if the dependence
of the temperature $T$ on the other thermodynamical quantities, in particular $\rho$ and $p$
is known.  For this reason we shall now make the following assumption about the matter system and its
(non-equilibrium!)  dynamics:

{\it The matter system is assumed to consist 
of $n_{\rm eff}$ species of effectively massless degrees of freedom
which all have the same temperature $T$. The equation of state is $p=\rho/3$, 
\ie $w=1/3$, and $\rho$ depends on $T$ as 
\be\label{3.9}
\rho(T) =\kappa^4 \; T^4 , \;\;\;\;\;\;\; \kappa\equiv (\pi^2 \; n_{\rm eff}/30)^{1/4}
\ee
No assumption is made about the relation $s=s(T)$.}

The first assumption, radiation dominance and equal temperature, is plausible since we shall find
that there is no significant entropy production any more once $H(t)$ has dropped substantially below
$\mp$, after the crossover from the NGFP to the GFP.

The second assumption, eq.(\ref{3.9}), amounts to the hypothesis formulated in the introduction. 
While entropy generation is a non-adiabatic process we assume, following Lima \cite{lima1},
that the non-adiabaticity is as small as possible. More precisely, the approximation is that the 
{\it equilibrium} relations among $\rho$, $p$, and $T$ are still valid in the
non-equilibrium situation of a cosmology with entropy production. In this sense, (\ref{3.9})
is the extrapolation of the standard relation (\ref{1.3a}) to a ``slightly non-adiabatic''
process.

Note that while we used (\ref{1.3c}) in relating $\clp(t)$ to the entropy production and also
postulated eq.(\ref{1.3a}), we do not assume the validity of the formula for
the entropy density, eq.(\ref{1.3b}), a priori. We shall see that the latter is an automatic
consequence of the cosmological equations. 

To make the picture as clear as possible we shall
neglect in the following all ordinary dissipative processes in the cosmological fluid.

Using $p=\rho/3$ and (\ref{3.9}) in (\ref{3.8}) the entropy production rate can be evaluated as follows:
\ba\label{3.20}
&\clp(t)=& \kappa\;  \Big [ a^3 \rho^{-1/4} \; \dot \rho + 4\; a^3 \; H \; \rho^{3/4} \Big ]\\[2mm]
& \; \; \; \; \; \;\;\;=& \frac{4}{3} \; \kappa \; \Big [ a^3\; \frac{d}{dt}(\rho^{3/4}) 
+ 3 \; \dot{a} \; a^2 \; \rho^{3/4} \Big ]\nonumber\\[2mm]
& \; \; \; \; \; \;\;\;=& \; \frac{4}{3} \; \kappa \Big [ \; a^3 \; \frac{d}{dt}(\rho^{3/4}) 
+ \rho^{3/4}  \frac{d}{dt}(a^3) \Big ]\nonumber
\ea
Remarkably, $\clp$ turns out to be a total time derivative:
\be\label{3.21}
\clp(t) = \frac{d}{dt} \; \Big [ \frac{4}{3} \; \kappa \; a^3 \; \rho^{3/4} \Big ]
\ee
Therefore we can immediately integrate (\ref{3.5}) and obtain 
\be\label{3.22}
S(t)=\frac{4}{3}\; \kappa \; a^3\; \rho^{3/4} +S_{\rm c}
\ee
or, in terms of the proper entropy density, 
\be\label{3.23}
s(t) = \frac{4}{3} \; \kappa\; \rho(t)^{3/4} +\frac{S_{\rm c}}{a(t)^3}
\ee
Here $\ssc$ is a constant of integration. In terms of $T$, using 
(\ref{3.9}) again, 
\be\label{3.24}
s(t) = \frac{2\pi^2 }{45} \; n_{\rm eff} \; T(t)^3 +\frac{S_{\rm c}}{a(t)^3}
\ee

The final result (\ref{3.24}) is very remarkable for at least two reasons. First, 
for $\ssc=0$, Eq.(\ref{3.24}) has exactly the form (\ref{1.3b}) which is valid for radiation
in equilibrium. Note that we did not postulate this relationship, only the $\rho(T)$--law
was assumed. The equilibrium formula $s\propto T^3$ was {\it derived} from the 
cosmological equations, \ie the modified conservation law. This result makes the hypothesis
``non-adiabatic, but as little as possible'' selfconsistent. 

Second, if $\lim_{t\rightarrow 0} \; a(t) \rho(t)^{1/4}=0$, which is  actually the
case for the most interesting class of cosmologies we shall find, then $S(t\rightarrow 0)=S_c$
by eq.(\ref{3.22}). As we mentioned in the introduction, the most plausible initial value
of $S$ is $S=0$ which means a vanishing constant of integration $S_c$ here. But then,
with $S_c=0$, (\ref{3.22}) tells us that the {\it entire} entropy carried by the 
massless degrees of freedom is due to the RG running. So it indeed seems to 
be true that the entropy of the CMBR photons we observe today is due to a coarse graining but,
unexpectedly, not a coarse graining of the matter degrees of freedom but rather of the
gravitational ones which determine the background spacetime the photons propagate on.

We close this section with various comments.
As for the interpretation of the function $\clp(t)$, let us remark that it also measures
the deviations from the classical laws $a^4\rho = const$ and $aT=const$, respectively,  since
we have $\clp = \frac{4}{3}\; \kappa \; d (a^4 \rho)^{3/4} /dt= \frac{4}{3}\kappa^4\;  {d}(aT)^3 /dt$.

Both in classical and in improved cosmology with the ``consistency condition'' imposed
the quantity $\clm \equiv 8\pi a^4\rho$ is conserved in time \cite{cosmo1}. If energy transfer
is permitted and the entropy of the ordinary matter grows, $\clm$ increases as well. This is obvious
from
\be\label{3.30}
\frac{d}{dt} \clm (t)^{3/4} = \frac{3}{4\kappa}(8\pi)^{3/4}\; \clp (t)
\ee
or, in integrated form, $S(t)=\frac{4}{3}\kappa (8\pi)^{-3/4} {\cal M}(t)^{3/4} +\ssc$ .

In a spatially flat Robertson-Walker spacetime the overall scale of $a(t)$ has no physical
significance. If $\clm$ is time independent, we can fix this gauge ambiguity by picking a 
specific value of $\clm$ and expressing $a(t)$ correspondingly. For instance, parametrized in this
way, the scale factor of the classical FRW cosmology with $\Lambda=0$, $w=1/3$ reads
\cite{cosmo1}
\be\label{3.31}
a(t)=  [ 4\barg \clm /3 ]^{1/4} \; \sqrt{t}
\ee
If, during the expansion, $\clm$ increases slowly, eq.(\ref{3.31}) tells us that the 
expansion is actually {\it faster} than estimated classically. Of course, what we actually 
have to do in order to  find the corrected $a(t)$ is to solve the improved field equations
and not insert $\clm=\clm(t)$ into the classical solution, in particular when the 
change of $\clm$ is not ``slow''. Nevertheless, this simple argument makes it clear that 
entropy production implies an increase of $\clm$ which in turns implies an extra increase
of the scale factor. This latter increase, or ``inflation'', is a pure quantum effect.
The explicit solutions to which we turn next will confirm this picture.

\section{Solving the RG improved Einstein Equations}
In \cite{ourfluc} we solved the improved Einstein equations (\ref{2.5a}, \ref{2.5b}) for the trajectory with 
realistic parameter values which was discussed in Section 5. The solutions were determined by applying
the algorithm described at the end of Section 4. Having fixed the RG trajectory, there exists 
a 1-parameter family of solutions $(H(t),\rho(t))$. This parameter is conveniently chosen to be the relative
vacuum energy density in the fixed point regime, $\OLS$. 

The very early part of the cosmology can be described analytically. For $k\rightarrow \infty$ the trajectory
approaches the NGFP, $(g,\lambda)\rightarrow (g_\ast,\lambda_\ast)$,  so that $G(k)=g_\ast/k^2$ and 
$\Lambda(k)=\lambda_\ast k^2$. In this case the differential equation can be solved analytically, with the result
\be\label{4.12}
H(t)=\alpha/t, \;\;\;\; a(t) = At^\alpha, \;\;\;\;\; \alpha = \Big [ \frac{1}{2}(3+3w)(1-\OLS) \Big]^{-1}
\ee
and  
\[
\rho(t)=\widehat\rho t^{-4}, \;\;\;\;\; G(t) = \widehat G  t^2,  \;\;\;\;  \Lambda(t) = \widehat\Lambda  /t^2.
\]
Here $A$, $\widehat\rho$, $\widehat G $, and $\widehat\Lambda $ are positive constants.
They depend on $\OLS$ which assumes
values in the interval $(0,1)$.
If $\alpha>1$ the deceleration parameter $q=\alpha^{-1} -1$ is  negative and the 
Universe is in a phase of {\it power law inflation}. Furthermore, it has 
{\it no particle horizon} if
 $\alpha \geq 1$, but does have a horizon of radius $d_{H}=t/(1-\alpha)$
if $\alpha < 1$. In the case of $w=1/3$ this means that there is  a horizon for 
$\OLS< 1/2$, but none if $\OLS \geq 1/2$.

If $w=1/3$, the above discussion of entropy generation applies. 
The corresponding production rate reads
\[
\clp(t) = 4 \kappa \; (\alpha-1) \; A^3\; \widehat{\rho}^{3/4} \; t^{3\alpha -4}.
\]
For the entropy per unit comoving volume we
find, if $\alpha \not =1$, 
\[
S(t) = \ssc + \frac{4}{3} \kappa \; A^3 \; \widehat{\rho}^{3/4}\; t^{3(\alpha -1)},
\]
and the corresponding proper entropy density is 
\[
s(t) =\frac{\ssc}{A^3 \; t^{3\alpha}}+ \frac{4\kappa \; \widehat\rho ^{3/4}}{3\;  t^3}.
\]
For the discussion of the entropy we must distinguish 3 qualitatively different cases.

\noindent
{\bf (a) The case $\mathbf{\alpha >1}$, \ie $\mathbf {1/2< \OLS <  1}$}: Here $\clp(t)>0$ so that the 
entropy and energy content of the matter system increases with time. By eq.(\ref{3.7}), 
$\clp >0$ implies $\dot\Lambda +8\pi  \rho  \dot G <0$. Since $\dot\Lambda <0$ but $\dot G>0$
in the NGFP regime, the energy exchange is predominantly due to the decrease of $\Lambda$
while the increase of $G$ is subdominant in this respect.

The comoving entropy $S(t)$ has a finite limit for $t\rightarrow 0$, 
$S(t\rightarrow 0) =\ssc$, and $S(t)$ grows monotonically for $t>0$. If $\ssc=0$,  
which would be the most
natural value in view of the discussion in the introduction, {\it all}  of the entropy carried
by the matter fields is due to the energy injection from $\Lambda$. 

\noindent
{\bf (b) The case $\mathbf{\alpha < 1}$, \ie $\mathbf{ 0<\OLS< 1/2}$}: 
Here $\clp(t)<0$ so that the energy and entropy of matter decreases. Since $\clp <0$ 
amounts to $\dot\Lambda +8\pi  \rho  \dot G>0$, the dominant physical effect is the increase of 
$G$ with time, the counteracting decrease of $\Lambda$ is less important.
The comoving entropy starts out from an infinitely positive value at the initial 
singularity, $S(t\rightarrow 0) \rightarrow +\infty$. This case is unphysical probably.

\noindent
{\bf (c) The case $\mathbf{\alpha=1}$,  $\mathbf{\OLS=1/2}$}: Here $\clp(t)\equiv 0$, $S(t)=const$.
The effect of a decreasing
$\Lambda$ and increasing $G$ cancel exactly.

At lower scales the RG trajectory leaves the NGFP and very rapidly ``crosses over'' to the GFP.
This is most clearly seen in the behavior of the anomalous dimension 
$\eta_{\rm N}(k)\equiv k\partial_k \ln G(k)$ which quickly changes from its NGFP value $\eta_\ast =-2$ 
to the classical $\eta_{\rm N}=0$. This transition happens near $k\approx \mp$ or, since $k(t)\approx H(t)$,
near a cosmological ``transition" time $t_{\rm tr}$ defined by the condition 
$k(t_{\rm tr})=\xi H(t_{\rm tr})=\mp$. (Recall that $\xi=O(1)$).
The complete solution to the improved equations can be found with numerical methods only. It 
proves convenient to use logarithmic variables normalized with respect to their respective values at the turning
point. Besides the ``RG time" 
$\tau \equiv \ln (k/\kat)$, we use $x \equiv \ln (a/\aT)$, 
$y \equiv \ln (t/\tT)$, and ${\cal U} \equiv \ln (H/\hT)$.

Summarizing the numerical results one can say  that for any value of $\OLS$ the UV cosmologies
consist of two scaling regimes and a relatively sharp crossover region near
$k,H\approx \mp$ corresponding to $x\approx -34.5$  
which connects them. At higher $k$-scales the fixed point approximation 
is valid, at lower scales one has a classical FRW cosmology in which $\Lambda$
can be neglected.

\begin{figure}[h]
\begin{center}
\includegraphics[width=6in]{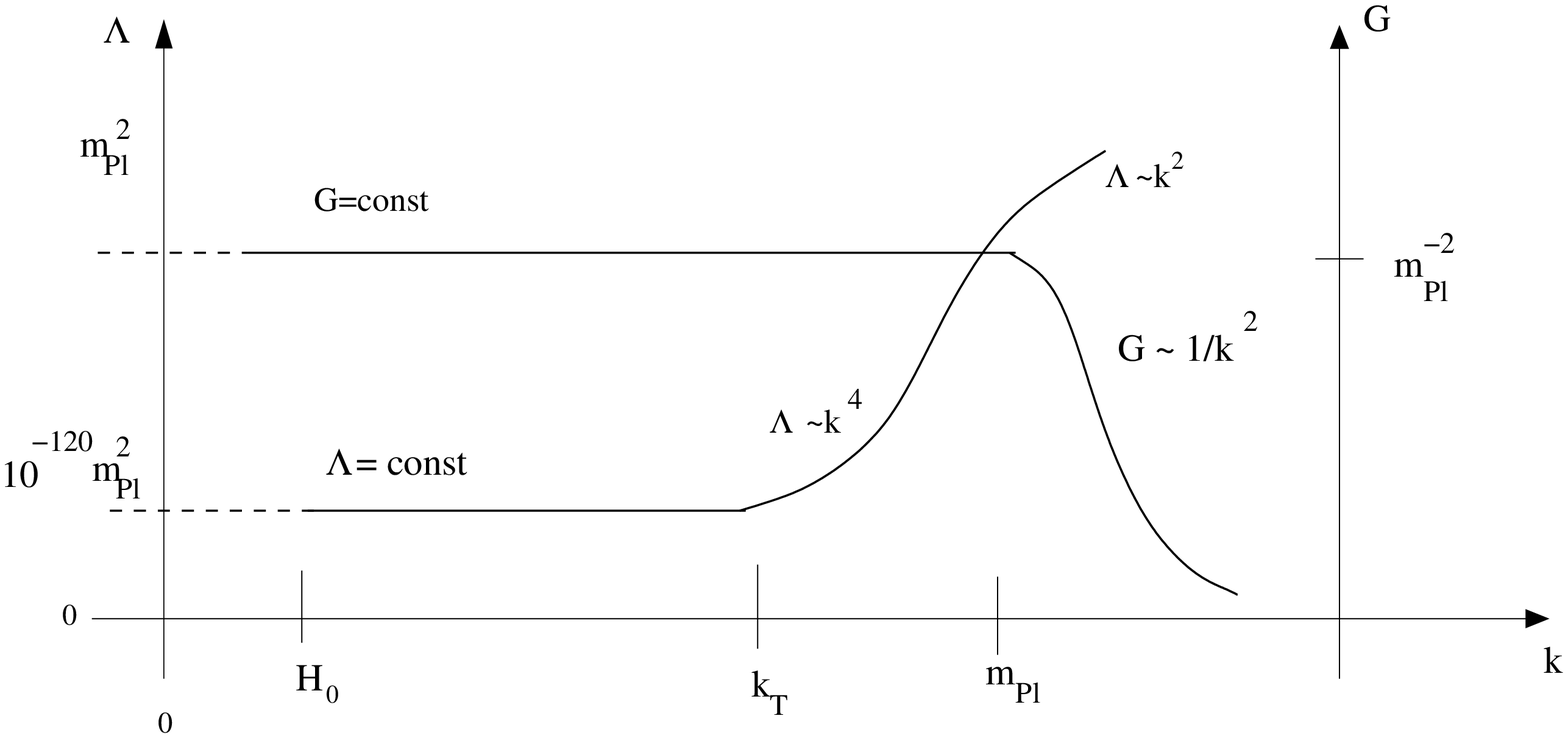}
\end{center}
\caption{The dimensionful quantities $\Lambda(k)$ and $G(k)$ for the RG trajectory with
realistic parameter values.}
\end{figure}

\begin{figure}[t]
\begin{center}
\includegraphics[width=15cm, height=11cm]{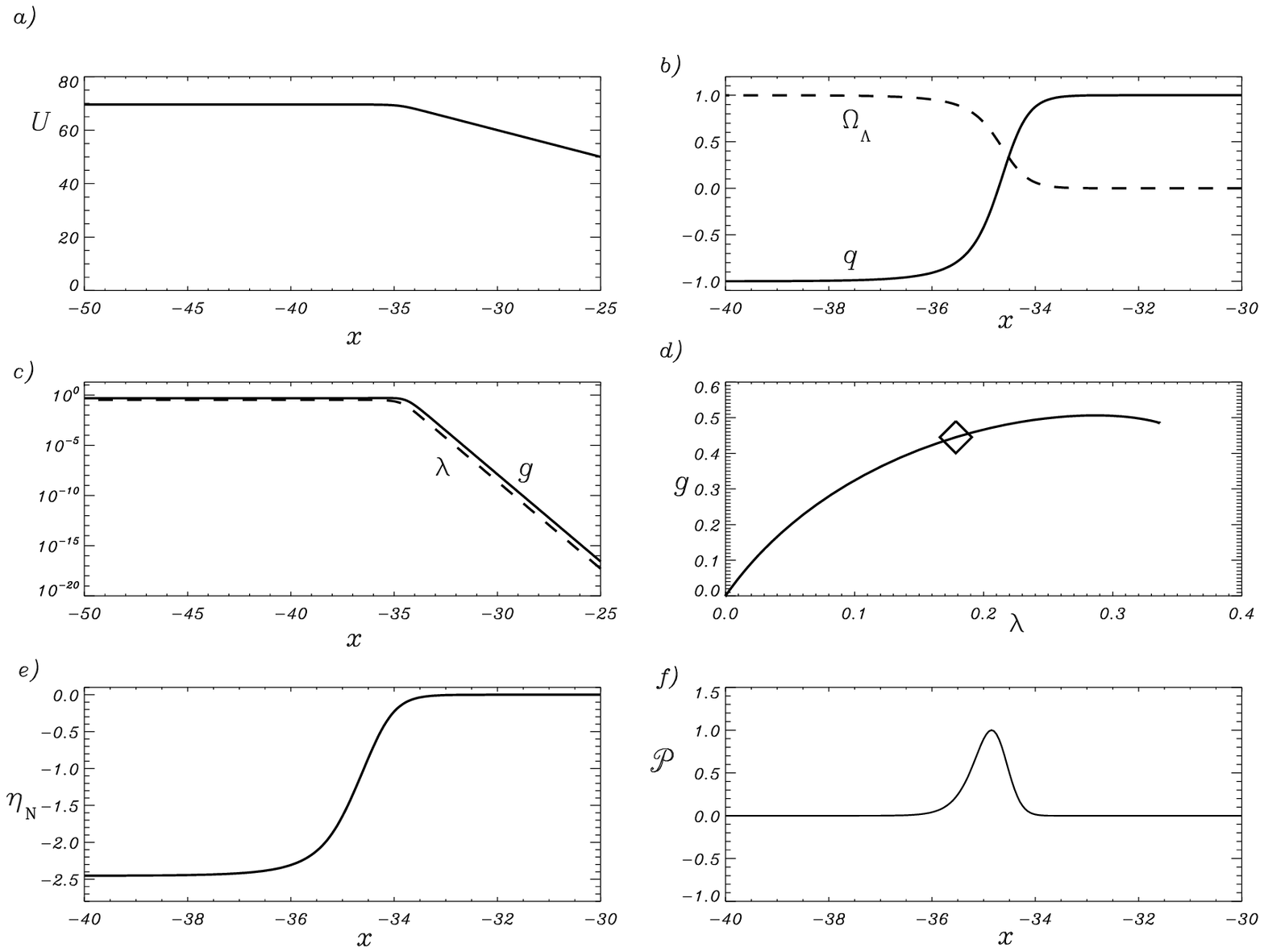}
\end{center}
\caption{The  crossover epoch of the cosmology for $\OLS=0.98$. The plots a), b), c) display the logarithmic Hubble
parameter $\UU$, as well as $q$, $\OL$, $g$ and $\lambda$ as a function of the logarithmic
scale factor $x$. A crossover is observed near $x\approx -34.5$. 
The diamond in plot d) indicates the point on the RG trajectory corresponding to this $x$-value. 
(The lower horizontal part of the trajectory is not visible on this scale.)
The plots e) and f) show the $x$-dependence of the anomalous dimension and entropy
production rate, respectively.}
\label{fig7}
\end{figure}
As an example, Fig.3 shows the crossover cosmology with  $\OLS=0.98$ and $w=1/3$. 
The entropy production rate $\clp$ is maximum at $t_{\rm tr}$ and quickly goes to zero for 
$t>t_{\rm tr}$; it is non-zero for all $t<t_{\rm tr}$. By varying the $\OLS$-value one can
check that the early cosmology is indeed described by the NGFP solution (5.1). For the logarithmic
$H$ vs. $a$- plot, for instance, it predicts $\UU=-2(1-\OLS)x $ for $ x<-34.4$. The left part of the plot in
Fig.3a and its counterparts with different values of $\OLS$ indeed comply with this relation.
If $\OLS \in (1/2,1)$ we have $\alpha = (2-2\OLS)^{-1}>1$ and $a(t)\propto t^\alpha$ describes a phase of
accelerated power law inflation. 

When $\OLS \nearrow 1$ the slope of $\UU (x) = -2 (1-\OLS)x$ decreases and
finally vanishes at $\OLS=1$. This limiting case corresponds  to a constant Hubble parameter, 
i.e. to de Sitter space. For values of $\OLS$ smaller than, but close to $1$ this 
de Sitter limit is approximated by an expansion $a\propto t^\alpha$ with a very large
exponent $\alpha$. 

The phase of power law inflation automatically comes to a halt 
once the RG running has reduced $\Lambda$ to a value where the resulting vacuum energy density no longer
can overwhelm the matter energy density.

\section{Inflation in the fixed point regime}
Next we discuss in more detail the epoch of power law inflation which is realized
in the NGFP regime if $\OLS > 1/2$. Since 
the transition from the fixed point to the classical FRW regime is rather sharp it will be
sufficient to approximate the RG improved UV cosmologies by the following caricature :
For $0<t< t_{\rm tr}$, the scale factor behaves as
$a(t)\propto t^\alpha$, $\alpha > 1$.  Here $\alpha = (2-2\OLS)^{-1}$ since $w =1/3$ 
will be assumed. Thereafter, for $t>t_{\rm tr}$, we have a classical, entirely
matter-driven expansion $a(t)\propto t^{1/2}$ . 
\subsection{Transition time and apparent initial singularity}
The transition time $t_{\rm tr}$ is dictated by the RG trajectory. 
It leaves the asymptotic scaling 
regime near $k\approx \mp$. Hence  $H(t_{\rm tr})\approx \mp$ and since $\xi=O(1)$ and $H(t)=\alpha/t$
we find the estimate
\be\label{6.2}
t_{\rm tr}= \alpha \; \tp
\ee
Here, as always, the Planck mass, time, and length are defined in terms of the value of Newton's
constant in the classical regime :
$\tp = \lp = \mp^{-1} = \bar{G}^{1/2} = G_{\rm observed}^{1/2}$.
Let us now assume that $\OLS$ is very close 
to $1$ so that $\alpha$ is large:
$\alpha \gg 1$. Then (\ref{6.2}) implies that the transition takes place at a cosmological time 
which is much later 
than the Planck time. At the transition the {\it Hubble parameter} is of order $\mp$, but the 
{\it cosmological time } is in general not of the order of $\tp$. Stated differently, the ``Planck time''
is {\it not} the time at which $H$ and the related physical quantities assume Planckian values. 
The Planck time as defined above is well within the NGFP regime:
$\tp = t_{\rm tr} / \alpha \ll t_{\rm tr}$. 

At $t=t_{\rm tr}$ the NGFP solution is to be matched continuously with a FRW cosmology
(with vanishing cosmological constant ). We may use the  classical formula $a\propto \sqrt{t}$ 
for the scale factor, but we must shift the time axis on the classical side such that $a$, 
$H$, and then as a result of (\ref{2.5a}) also $\rho$ are continuous at $t_{\rm tr}$. Therefore
$a(t)\propto (t-t_{\rm as})^{1/2}$ and 
$H(t) = \frac{1}{2} \; (t-t_{\rm as})^{-1} \;\;\; \text{for } \;\;\; t> t_{\rm tr}$.
Equating this Hubble parameter  at $t=t_{\rm tr}$ to 
$H(t) = \alpha /t$, valid in the NGFP regime, we find that the shift $t_{\rm as}$ must be chosen
as 
\[
t_{\rm as} =  (\alpha -\frac{1}{2}) \tp = (1 - \frac{1}{2\alpha})  t_{\rm tr}  <  t_{\rm tr}. 
\]
Here the subscript 'as' stands for ``apparent singularity''. This is to indicate that if one continues
the classical cosmology to times $t<t_{\rm tr}$, it has an initial singularity (``big bang'') at 
$t=t_{\rm as}$. Since, however, the FRW solution is not valid there nothing special happens at 
$t_{\rm as}$; the true initial singularity is located at $t=0$ in the NGFP regime. (See Fig.~4.)
\subsection{Crossing the Hubble radius}
In the NGFP regime $0<t< t_{\rm tr}$ the Hubble radius $\ell_{H} (t) \equiv 1/H(t)$, i.e.
$\ell_{H} (t) = t/{\alpha} $ , 
increases linearly with time but, for $\alpha \gg 1$, with a very small slope. At the transition, the slope 
jumps from $1/\alpha$ to the value $2$ since $H=1/(2t)$ 
and $\ell_{H}=2t$ in the  FRW regime. This behavior is sketched in Fig.~4. 

Let us consider some structure of comoving length $\Delta x$, 
a single wavelength of a density perturbation,
for instance. The corresponding physical, i.e. proper length is
\[
L(t) = a(t) \Delta x  
\]
then.
In the NGFP regime it has the time dependence 
$L(t) =  ({t}/{\tr}  )^{\alpha} \; L(\tr)$.
The ratio of $L(t)$ and the Hubble radius evolves according to 
\[
\frac{L(t)}{\LH (t)} =  ( \frac{t}{\tr}  )^{\alpha-1} \; \frac{L(\tr)}{\LH (\tr)}.
\]
For $\alpha > 1$, i.e. $\OLS > 1/2$, the proper length of any object grows faster than the Hubble
radius. So objects which are of ``sub-Hubble'' size at early times can cross the Hubble radius and become
``super-Hubble'' at later times, see Fig.~4. 

Let us focus on a structure which, at $t=\tr$, is
$e^N$ times larger than the Hubble radius. Before the transition we have
$L(t)/\LH (t) = e^N \; (t/\tr)^{\alpha -1}$.
Assuming $e^N > 1$, there exists a time $t_N < \tr$ at which $L(t_{N}) =\LH (t_{ N})$
so that the structure considered ``crosses'' the Hubble radius at the time $t_N$. It is given
by 
\be\label{6.8}
t_{N}=\tr \;{\rm exp} {\Big ( -\frac{N}{\alpha -1} \Big )} 
\ee
What is remarkable about this result is that, even with rather moderate values of $\alpha$, one can 
easily ``inflate'' structures to a size which is by many $e$-folds larger than the Hubble radius 
{\it during a very short time interval at the end of the NGFP epoch}. 

Let us illustrate this phenomenon by means of an example, namely the choice $\OLS = 0.98$ used
in Fig.~3.
Corresponding to $98\%$ vacuum and $2\%$ matter energy density in the NGFP regime, this value
is  still ``generic'' in the sense that $\OLS$ is not fine tuned to equal unity 
with a precision of many decimal places. It leads to the exponent $\alpha = 25$, the transition
time $\tr = 25 \; \tp$, and $t_{\rm as}=24.5 \; \tp$. 

The largest structures in the present Universe, evolved backward in time by the classical equations
to the point where $H=\mp$, have a size of about $e^{60}\; \lp$ there. We can use 
(\ref{6.8}) with $N=60$ to find the time $t_{\rm 60}$ at which those structures crossed
the Hubble radius. With $\alpha = 25$ the result is $t_{\rm 60}=2.05\; \tp = \tr /12.2$. Remarkably, $t_{\rm 60}$ is smaller than 
$\tr$ by one order of magnitude only. As a consequence, the physical conditions prevailing at the
time of the crossing are not overly  ``exotic'' yet. 
The Hubble parameter, for instance, is only one order of magnitude larger than at the transition:
$H(t_{\rm 60})\approx 12 \mp$.  The same is true for the temperature; one can show that 
$T(t_{\rm 60})\approx 12 T(\tr)$ where $T(\tr)$ is of the order of $\mp$. Note  that $t_{\rm 60}$ is larger than $\tp$.
\subsection{Primordial density fluctuations}
\begin{figure}[t]
\begin{center}
\includegraphics[width=10cm]{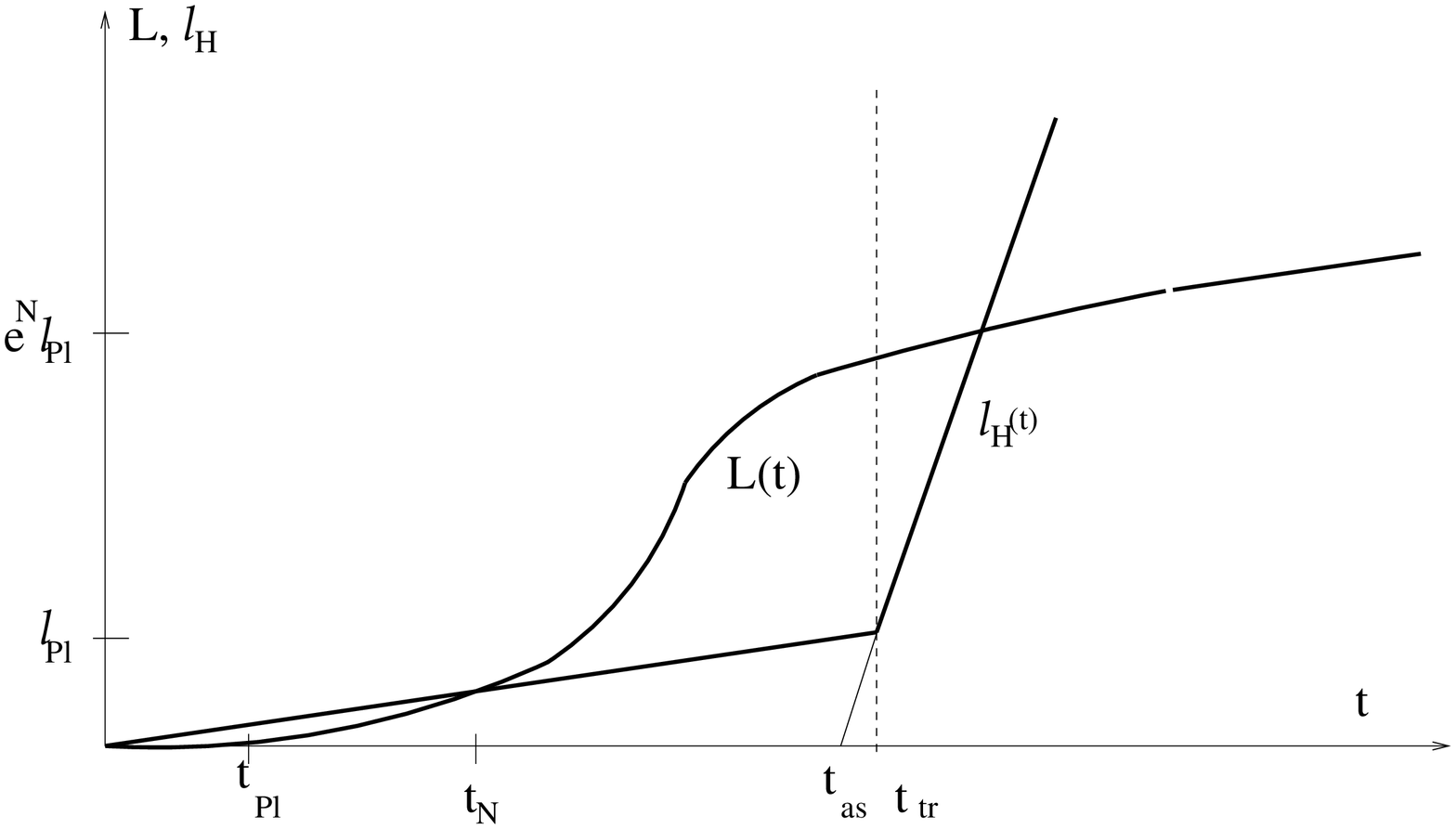}
\end{center}
\caption{Shown is the proper length $L$ and the Hubble radius as a function of time. 
The NGFP and FRW cosmologies are valid for $t<t_{\rm tr}$ and $t>t_{\rm tr}$, respectively.
The classical cosmology has an apparent initial singularity at $t_{as}$ outside its domain of 
validity. Structures of size $e^N \lp$ at $t_{\rm tr}$ cross the Hubble radius at $t_N$, a 
time which can be larger than the Planck time. }
\label{fig9}
\end{figure}
QEG offers a natural mechanism for generating primordial fluctuations during the NGFP epoch.  They  have a
scale free spectrum with a spectral index close to $n=1$. This mechanism is at the very heart of the 
Asymptotic Safety underlying the nonperturbative renormalizability of QEG. 
A detailed discussion of this mechanism is beyond the scope of the present review; the reader it referred to 
\cite{oliver2,cosmo1,ourfluc}. Suffice it to say that the quantum mechanical generation of the primordial fluctuations
happens on sub-Hubble distance scales. However, thanks to the inflationary NGFP era the modes relevant
to cosmological structure formation were indeed smaller than the Hubble radius at a sufficiently early time,
for $t<t_{\rm 60}$, say. (See the $L(t)$ curve in Fig.4.)

\section{Conclusions}
We advocated the point of view that the scale dependence of the gravitational
parameters has an impact on the physics of the universe we live in and we tried to identify known
features of the Universe which could possibly  be due to this scale dependence. We proposed three possible
candidates for such features: the entropy carried by the radiation which fills the Universe today,
a period of automatic, $\Lambda$-driven inflation that requires no ad hoc inflaton, and the primordial
density perturbations. While there is clearly no direct observational evidence for inflation it can 
explain  super-Hubble sized perturbations. For further details we refer to \cite{ourfluc} while for an 
extension of this approach including higher order operators in the truncated action we 
refer to \cite{bcp}.



\bibliographystyle{mdpi}
\makeatletter
\renewcommand\@biblabel[1]{#1. }
\makeatother

\begin{thebibliography}{1}
\bibitem{mr}
M.~Reuter,
Phys.\ Rev.\ D 57 (1998) 971 and hep-th/9605030.
%
\bibitem{percadou}
D.~Dou and R.~Percacci,
Class.\ Quant.\ Grav.\ 15 (1998) 3449.
%
\bibitem{oliver1}
O.~Lauscher and M.~Reuter,
Phys.\ Rev.\ D 65 (2002) 025013 and hep-th/0108040.
%
\bibitem{frank1}
M.~Reuter and F.~Saueressig,
Phys.\ Rev.\ D 65 (2002) 065016 and hep-th/0110054.
%
\bibitem{oliver2}
O.~Lauscher and M.~Reuter, Phys.\ Rev.\ D 66 (2002) 025026 and
 hep-th/0205062.
%
\bibitem{oliver3}
O.~Lauscher and M.~Reuter,
Class.\ Quant.\ Grav.\ 19 (2002) 483 and hep-th/0110021.
%
\bibitem{oliver4}
O.~Lauscher and M.~Reuter,
Int.\ J.\ Mod.\ Phys.\ A 17 (2002) 993 and hep-th/0112089.
%
\bibitem{frank2}
M.~Reuter and F.~Saueressig,
Phys.\ Rev.\ D 66 (2002) 125001 and hep-th/0206145;
Fortschr.\ Phys.\ 52 (2004) 650 and hep-th/0311056.
%
\bibitem{souma}
W.~Souma,
Prog.\ Theor.\ Phys.\ 102 (1999) 181.
%
\bibitem{perper1}
R.~Percacci and D.~Perini,
Phys.\ Rev.\ D 67 (2003) 081503;
Phys.\ Rev.\ D 68 (2003) 044018;
D.~Perini, Nucl.\ Phys.\ Proc.\ Suppl.\ 127 C (2004) 185.
%
\bibitem{codello}
A.~Codello and R.~Percacci, Phys.\ Rev.\ Lett. 97 (2006) 221301;
A.~Codello, R.~Percacci and C.~Rahmede, hep-th/0705.1769 ;
P.~Machado and F.~Saueressig, arXiv:\ 0712.0445 \ [hep-th].
%
\bibitem{litimgrav}
D.~Litim, Phys.\ Rev.\ Lett.\ 92 (2004) 201301; 
P.~Fischer and D.~Litim, Phys. Lett. B 638  (2006) 497.
%
\bibitem{prop}
A.~Bonanno and M.~Reuter, JHEP\ 02 (2005) 035 and hep-th/0410191.
\bibitem{colombia} For a review see: M.~Reuter and F.~Saueressig, arXiv: \ 0708.1317 \ [hep-th].
%
%
\bibitem{essential}
R.~Percacci and D.~Perini,
Class.\ Quant.\ Grav. 21 (2004) 5035
and hep-th/0401071.
%
%
\bibitem{wein}
S.~Weinberg
in \textit{General Relativity, an Einstein Centenary Survey},
S.W.~Hawking and W.~Israel (Eds.),
Cambridge University Press (1979);
S.~Weinberg, hep-th/9702027, arXiv:\ 0903.0568, arXiv:\ 0908.1964.

\bibitem{livrev}
For a review see: M.~Niedermaier and M.~Reuter, Living Reviews in Relativity,  9 (2006) 5.

\bibitem{max}
P.~Forg\'acs and M.~Niedermaier,
hep-th/0207028;
M.~Niedermaier, JHEP 12 (2002) 066;
Nucl.\ Phys.\ B 673 (2003) 131;
preprint gr-qc/0610018.
%
\bibitem{avact}
C.~Wetterich,
Phys.\ Lett.\ B 301 (1993) 90.
%
\bibitem{ym}
M.~Reuter and C.~Wetterich,
Nucl.\ Phys.\ B 417 (1994) 181,
Nucl.\ Phys.\ B 427 (1994) 291,
Nucl.\ Phys.\ B 391 (1993) 147,
Nucl.\ Phys.\ B 408 (1993) 91;
M.~Reuter,
Phys.\ Rev. D 53 (1996) 4430,
Mod.\ Phys.\ Lett.\ A 12 (1997) 2777.
%
\bibitem{avactrev}
For a review see:
J.~Berges, N.~Tetradis and C.~Wetterich,
Phys.\ Rep.\ 363 (2002) 223;
C.~Wetterich,
Int.\ J.\ Mod.\ Phys.\ A 16 (2001) 1951;
H.~Gies, hep-ph/0611146.
\bibitem{bh}
A.~Bonanno and M.~Reuter,
Phys.\ Rev.\ D 62 (2000) 043008 and  hep-th/0002196;
Phys.\ Rev.\ D 73 (2006) 083005 and hep-th/0602159;
Phys.\ Rev.\ D 60 (1999) 084011 and gr-qc/9811026.
%
\bibitem{erik1} 
M.~Reuter and E.~Tuiran, hep-th/0612037.
%
\bibitem{cosmo1}
A.~Bonanno and M.~Reuter,
Phys.\ Rev.\ D 65 (2002) 043508 and hep-th/0106133.
%
\bibitem{cosmofrank}
M.~Reuter and F.~Saueressig, JCAP\ 09 (2005) 012 and hep-th/0507167.
%
\bibitem{ourfluc} A.~Bonanno and M.~Reuter, JCAP 08(2007) 024 and arXiv:\ 0706.0174\ [hep-th].
\bibitem{pune} A.~Bonanno and M.~Reuter, Journal of Phys. Conf. Ser. 140(2008)   012008 and arXiv:\ 0803.2546\ [hep-th].
\bibitem{weinf} S.~Weinberg, Phys.\ Rev.\ D 81 (2010) 083535 and  arXiv:\ 0911.3165.
%
\bibitem{cosmo2}
A.~Bonanno and M.~Reuter,
Phys.\ Lett.\ B 527 (2002) 9 and astro-ph/0106468;
Int.\ J.\ Mod.\ Phys.\ D 13 (2004) 107 and astro-ph/0210472.
%
\bibitem{elo}
E.~Bentivegna, A.~Bonanno and M.~Reuter,
JCAP 01 (2004) 001  and
astro-ph/0303150.
\bibitem{esposito}
A.~Bonanno, G.~Esposito and C.~Rubano,
Gen.\ Rel.\ Grav.\ 35 (2003) 1899;
Class.\ Quant.\ Grav.\ 21 (2004) 5005;
A.~Bonanno, G.~Esposito, C.~Rubano and P.~Scudellaro,
Class.\ Quant.\ Grav. 23 (2006) 3103;  preprint gr-qc/0610012.
%
\bibitem{h1}
M.~Reuter and H.~Weyer,
Phys.\ Rev.\ D 69 (2004) 104022
and hep-th/0311196;
M.~Reuter and H.~Weyer,
Phys.\ Rev.\ D 70 (2004) 124028
and  hep-th/0410117;
M.~Reuter and H.~Weyer,
JCAP\ 12 (2004) 001
and hep-th/0410119.
%
\bibitem{lima1} J.A.S.~Lima, Phys.~Rev. D54 (1996) 2571.
\bibitem{lima2} J.A.S.~Lima,  Gen.~Rel.~Grav. 29 (1997) 805.
\bibitem{bcp}
A.~Bonanno, A.~Contillo, R.~Percacci, arXiv:\ 1006.0192.



\end{thebibliography}

\end{document}